%% file: main.tex
\documentclass[sigconf]{acmart}

\input{preamble/packages}

\input{preamble/definitions}

\input{preamble/authors}
\input{preamble/metadata}

\title[FairDiverse: A Comprehensive Toolkit for Fair and Diverse Information Retrieval Algorithms]{FairDiverse: A Comprehensive Toolkit for Fair and Diverse \\Information Retrieval Algorithms}

\begin{document}

\begin{abstract}
In modern information retrieval (IR), achieving more than just accuracy is essential to sustaining a healthy ecosystem, especially when addressing fairness and diversity considerations. To meet these needs, various datasets, algorithms, and evaluation frameworks have been introduced. However, these algorithms are often tested across diverse metrics, datasets, and experimental setups, leading to inconsistencies and difficulties in direct comparisons.
This highlights the need for a comprehensive IR toolkit that enables standardized evaluation of fairness- and diversity-aware algorithms across different IR tasks. To address this challenge, we present \textbf{FairDiverse}, an open-source and standardized toolkit.
FairDiverse offers a framework for integrating fairness- and diversity-focused methods, including pre-processing, in-processing, and post-processing techniques, at different stages of the IR pipeline. The toolkit supports the evaluation of \textbf{28} fairness and diversity algorithms across \textbf{16} base models, covering two core IR tasks—search and recommendation—thereby establishing a comprehensive benchmark.
Moreover, FairDiverse is highly extensible, providing multiple APIs that empower IR researchers to swiftly develop and evaluate their own fairness- and diversity-aware models, while ensuring fair comparisons with existing baselines. The project is open-sourced and available on GitHub:~\url{https://github.com/XuChen0427/FairDiverse}.

\end{abstract}

\maketitle

\section{Introduction}


Information retrieval (IR) tasks, such as search and recommendation, typically aim to select the information that meets user needs~\cite{IRbook, chowdhury2010introduction}. 
In modern IR, factors beyond the accuracy of information access, such as novelty, diversity, and fairness, are crucial for building a healthy ecosystem~\cite{Li_sigir24}. 
Among these factors, fairness and diversity have gained increasing attention in recent years~\cite{li2022fairness, santos2010exploiting, PM2_12_sigir}. Both aim to expose users to a broader range of information sources~\cite{santos2010exploiting, dang2012diversity} while also supporting diverse types of providers~\cite{fairrec, xu2023p}.

\begin{figure*}[t]  
    \centering    
    \includegraphics[width=\linewidth]{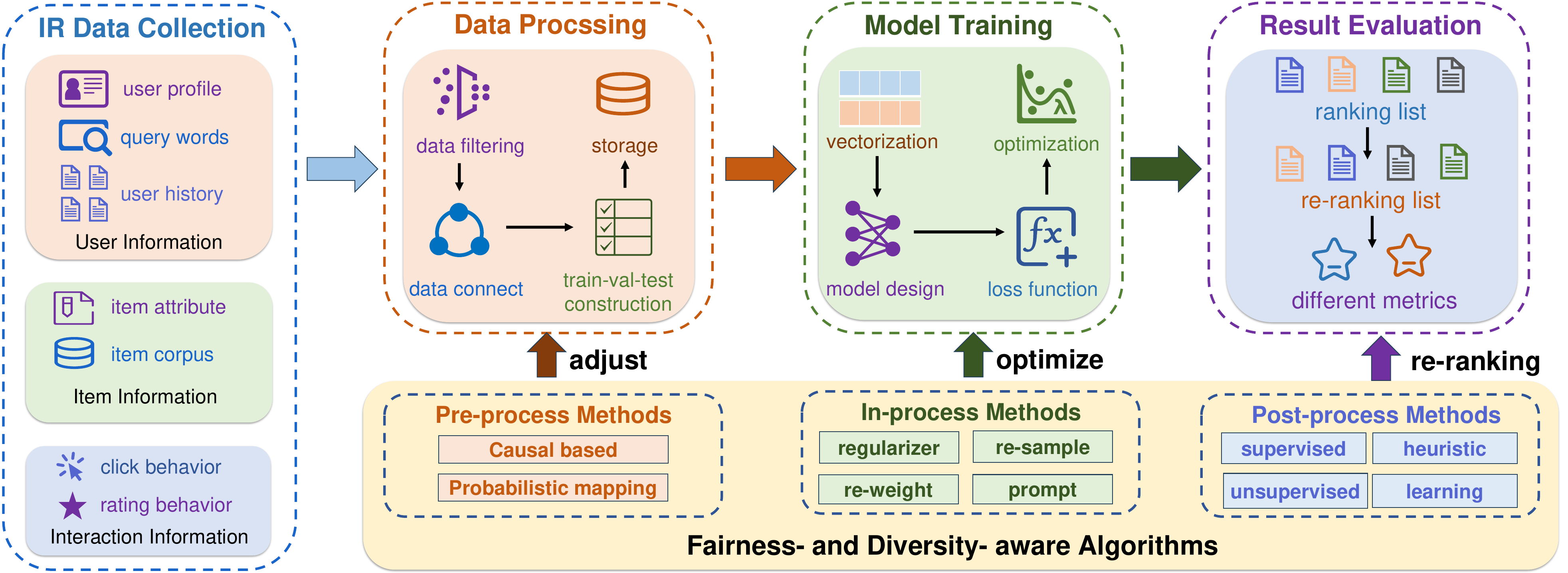}
    \caption{Overall architecture of FairDiverse. We categorize fairness- and diversity-aware algorithms into pre-processing, in-processing, and post-processing stages, corresponding to data processing, model training, and result evaluation phases of IR.}
    \label{fig:pipline}
    \vspace*{-2mm}
\end{figure*}

\begin{table}[t]
\centering
\setlength{\tabcolsep}{1.1pt}
\caption{Comparison between existing fairness- and diversity-aware toolkits. \ding{55} denotes that the feature is not supported, while \ding{51} indicates that the feature is supported. }
\label{tab:compare}
\begin{tabular}{l cccccc }
\toprule
Features & 
\rotatebox{65}{Recbole \cite{recbole2.0}} & 
\rotatebox{65}{FFB \cite{han2023ffb}} & 
\rotatebox{65}{Fairlearn \cite{bird2020fairlearn}} & 
\rotatebox{65}{AIF360 \cite{aif360-oct-2018}} & 
\rotatebox{65}{Aequitas \cite{jesus2024aequitas}} & 
\rotatebox{65}{\textbf{FairDiverse}} \\ 
\midrule
Recommendation & \ding{51} & \ding{55} & \ding{55} & \ding{55} & \ding{55} & \ding{51} \\
Search & \ding{55} & \ding{55} & \ding{55} & \ding{55} & \ding{55} & \ding{51} \\
\midrule
Pre-processing & \ding{55} & \ding{55} & \ding{51} & \ding{51} & \ding{51} & \ding{51} \\
In-processing & \ding{51} & \ding{51} & \ding{51} & \ding{51} & \ding{51} & \ding{51} \\
Post-processing & \ding{55} & \ding{55} & \ding{51} & \ding{51} & \ding{51} & \ding{51} \\
\midrule
Number of models & 4 & 6 & 6 & 15 & 10 & \textbf{28} \\ \bottomrule
\end{tabular}
\vspace*{-3mm}
\end{table}

To ensure fairness and diversity in IR systems, many fairness-aware~\cite{fairrec, xu2023p, TaxRank, SDRO, APR, FairNeg, cpfair, rus2024study} and diversity-aware algorithms~\cite{li2022fairness, santos2010exploiting, qin2020diversifying, yan2021diversification} have been designed as plugins or modules that can be integrated into various stages of the IR pipeline. However, fairness and diversity often suffer from a lack of unified definitions~\cite{li2022fairness, LLM4FairSurvey}. 
As a result, the evaluation of these algorithms in IR systems are based on different metrics, datasets, and evaluation settings (details are shown in Section~\ref{sec:related_work}). Hence, the performance of these algorithms cannot be compared consistently. Developing a unified, fair, and extensible toolkit for fairness and diversity is critically important and urgently needed to evaluate these algorithms consistently across IR tasks. Such a toolkit framework holds significant value for fostering a trustworthy IR community.

To create a unified and equitable evaluation, we introduce FairDiverse, an open-source standardized toolkit designed to assess fairness and diversity in IR systems.
First, FairDiverse offers detailed guidance on incorporating fairness- and diversity-aware algorithms throughout various stages of the IR process. These algorithms are categorized into pre-processing, in-processing, and post-processing methods, corresponding to data processing, model training, and result evaluation stages in different IR pipeline steps, respectively. 
Then, FairDiverse implements a wide range of fairness- and diversity-aware models (\textbf{28} models) tailored to \textbf{16} base models under two fundamental IR tasks: search and recommendation. 
It offers corresponding implementation code and systematically evaluates these algorithms using over ten accuracy, fairness, and diversity metrics, enabling the construction of a benchmark within FairDiverse.


In the literature, only a few open-source toolkits and libraries have been developed for fairness- and diversity-aware IR algorithms. Table~\ref{tab:compare} provides a comparison between these existing resources and the proposed FairDiverse, highlighting features such as supported IR tasks (recommendation and search), algorithm types (pre-processing, in-processing, and post-processing), and the number of implemented models. Other toolkit details are provided in Section~\ref{sec:related_work}. As shown in Table~\ref{tab:compare}, FairDiverse provides the largest number of models, offering extensive coverage of all types of fairness- and diversity-aware algorithms. Additionally, it supports major information retrieval (IR) tasks, including search and recommendation, making it a versatile and comprehensive toolkit.

FairDiverse is designed to be highly extensible, providing a range of flexible APIs that allow IR researchers to efficiently develop and integrate their own fairness- and diversity-aware IR models. This extensibility ensures that researchers can tailor the toolkit to their specific needs while maintaining consistency with established evaluation protocols. This makes it an invaluable resource for advancing fairness and diversity in IR systems.

\input{sections/Pipeline}
\input{sections/details}

\input{sections/usage}
\input{sections/experiments}
\input{sections/APIs}
\input{sections/related_work}
\input{sections/Conclusion}

\newpage
\bibliographystyle{ACM-Reference-Format}
\bibliography{references}

\end{document}

%% file: preamble/packages.tex
\usepackage{hyperref}
\usepackage{latexsym}
\usepackage{mathrsfs}
\usepackage{multirow}
\usepackage{amsmath}
\usepackage{amsfonts}
\usepackage{subfigure}
\usepackage{graphicx}
\usepackage{tabularx}
\usepackage[boxed,ruled,lined]{algorithm2e}
\usepackage{algorithmic}
\usepackage{comment}
\usepackage{balance}
\usepackage{bm}
\usepackage[inline]{enumitem}
\usepackage[skip=1mm]{caption}

\usepackage{acronym}
\usepackage{pifont}
\usepackage{makecell}
\usepackage{listings}

\usepackage{xcolor}

\lstdefinestyle{shell}{
    language=bash,        
    basicstyle=\ttfamily, 
    keywordstyle=\color{black},   
    stringstyle=\color{blue},   
    commentstyle=\color{gray},   
    backgroundcolor=\color{black!5}, 
    breaklines=true,      
    frame=single,         
    showstringspaces=false, 
    morekeywords={python},
}

%% file: preamble/definitions.tex
\newcommand{\ie}{\emph{i.e., }}

\acrodef{IR}{information retrieval}

%% file: preamble/authors.tex
\author{Chen Xu}
\authornotemark[1]
\affiliation{%
  \institution{Renmin University of China}
  \city{Beijing}
  \country{China}
}
\email{xc\_chen@ruc.edu.cn}

\author{Zhirui Deng}
\authornotemark[1]
\affiliation{%
  \institution{Renmin University of China}
  \city{Beijing}
  \country{China}
}
\email{zrdeng@ruc.edu.cn}

\author{Clara Rus}
\authornotemark[1]
\affiliation{%
  \institution{University of Amsterdam}
  \city{Amsterdam}
  \country{The Netherlands}
}
\email{c.a.rus@uva.nl}

\author{Xiaopeng Ye}
\affiliation{%
  \institution{Renmin University of China}
  \city{Beijing}
  \country{China}
}
\email{xpye@ruc.edu.cn}

\author{Yuanna Liu}
\affiliation{%
  \institution{University of Amsterdam}
  \city{Amsterdam}
  \country{The Netherlands}
}
\email{y.liu8@uva.nl}

\author{Jun Xu}
\authornotemark[2]
\affiliation{%
    \institution{Renmin University of China}
  \city{Beijing}
  \country{China}
}
\email{junxu@ruc.edu.cn}

\author{Zhicheng Dou}
\authornotemark[2]
\affiliation{%
    \institution{Renmin University of China}
  \city{Beijing}
  \country{China}
}
\email{dou@ruc.edu.cn}

\author{Ji-Rong Wen}
\affiliation{%
    \institution{Renmin University of China}
  \city{Beijing}
  \country{China}
}
\email{jrwen@ruc.edu.cn}

\author{Maarten de Rijke}
\affiliation{%
  \institution{University of Amsterdam}
  \city{Amsterdam}
  \country{The Netherlands}
}
\email{m.derijke@uva.nl}

\thanks{$*$: Equal contributions. \dag: Jun Xu and Zhicheng Dou are the corresponding authors.}
\renewcommand{\shortauthors}{Chen Xu et al.}

%% file: preamble/metadata.tex
\copyrightyear{2025}
\acmYear{2025}
\setcopyright{rightsretained}

\acmConference[SIGIR '25]{The 48th International ACM SIGIR Conference on Research and Development in Information Retrieval}{July 13--18, 2025}{Padua, Italy}

\acmPrice{}
\acmDOI{}
\acmISBN{}

\ccsdesc[500]{Information systems~Information retrieval}

\keywords{Information Retrieval, Fairness, Diversity, Toolkit}

%% file: sections/Pipeline.tex
\vspace*{-2mm}
\section{Toolkit Overview}

In this section, we illustrate the overview pipelines of FairDiverse, as shown in Figure~\ref{fig:pipline}. Generally, search and recommendation tasks in IR can be considered ranking tasks with similar pipelines~\cite{yao2021user, xie2024unifiedssr}. Next, we will detail the IR pipeline steps, incorporating fairness- and diversity-aware algorithms.



\begin{table*}[t]
\setlength{\tabcolsep}{1.1pt}
\caption{The models implemented in FairDiverse. }
\label{tab:models}
\resizebox{\textwidth}{!}{%
\begin{tabular}{ll}
\toprule
Model types & Models \\ \hline
\multicolumn{2}{c}{\emph{Recommendation}} \\
\midrule
Base model & MF~\cite{DMF}, BPR~\cite{BPR}, GRU4Rec~\cite{tan2016improved}, SASRec~\cite{SASRec}, Llama3~\cite{dubey2024llama}, Qwen2~\cite{bai2023qwen}, Mistral~\cite{jiang2023mistral7b} \\
In-processing & APR~\cite{APR}, DPR~\cite{DPR}, FairDual~\cite{FairDual}, FairNeg~\cite{FairNeg}, FOCF~\cite{yao2017beyond}, IPS~\cite{jiang2024item}, Reg~\cite{Reg}, Minmax-SGD~\cite{Minmax-SGD}, SDRO~\cite{SDRO}, FairPrompts~\cite{xu-etal-2024-study} \\
Post-processing & P-MMF~\cite{xu2023p}, CP-Fair~\cite{cpfair}, FairRec~\cite{fairrec}, FairRec+~\cite{fairrecplus}, FairSync~\cite{fairsync}, min-regularizer~\cite{xu2023p}, Tax-Rank~\cite{TaxRank}, Welf~\cite{nips21welf}, RAIF~\cite{liu2025repeat} \\
\midrule
\multicolumn{2}{c}{\emph{Search}} \\
\midrule
Base model & MART \cite{friedman2001greedy}, RankNet \cite{burges2005learning}, RankBoost \cite{freund2003efficient}, AdaRank \cite{xu2007adarank}, Coordinate Ascent \cite{metzler2007linear}, LambdaMART \cite{wu2010adapting}, ListNet \cite{cao2007learning}, Random Forests \cite{breiman2001random} \\
Pre-processing & CIF-Rank~\cite{yang2020causal}, LFR~\cite{zemel2013learning}, gFair, iFair~\cite{lahoti2019ifair} \\
Post-processing & PM2~\cite{PM2_12_sigir}, xQuAD~\cite{xQuAD_10_www}, DESA~\cite{DESA_20_cikm}, DALETOR~\cite{DALETOR_21_WWW}, DiversePrompts (based on GPT-4o~\cite{openai2024gpt4technicalreport} and Claude 3.5~\cite{Claude3.5}) \\ 
\bottomrule
\end{tabular}%
}
\end{table*}

\noindent\textbf{IR data collection.}
First, we collect the user and item information. Let $\mathcal{U}$ denote the set of users, and $\mathcal{I}$ the set of items. Each user $u\in\mathcal{U}$ may have a different user profile $\mathcal{P}_u$ such as age, gender, occupation, etc. We will record the user's browsing historical item list $H_u=[i_1, i_2, \cdots, i_n]$.
Each item $i\in\mathcal{I}$ is associated with specific attributes, such as categories, descriptions, and other metadata.

When a user $u$ interacts with an IR system, they may actively input a query $q_u$ to explicitly express their information need, a scenario commonly referred to as a search task. Alternatively, the user may not provide a query; instead, they rely on the IR system to infer their information needs and deliver relevant content, which characterizes a recommendation task.

Meanwhile, the collected data should also capture user behaviors, such as clicks, ratings, and other interactions.
For example, click behavior $c_{u,i}=1$ indicates that the user has clicked on the item, while $c_{u,i}=0$ signifies that the user did not interact with the item on the browser or recommender platform. Rating behaviors $r\in [0,5]$ denotes the preference degree of the 
These interaction behaviors are typically regarded as labels to train the IR models.

\noindent\textbf{Data processing.}
After collecting the IR data, it is essential to preprocess the dataset by filtering out noisy data samples, such as removing users with very few interaction histories, to ensure the quality of the data~\cite{recbole}. Then, we integrate user and item information with interaction behaviors and partition the data into training, validation, and test sets for model training and evaluation.

The pre-processing algorithms are primarily applied at this stage, aiming to mitigate biases present in the model input before training~\cite{rus2024study}. Specifically, certain features may enhance model performance but are influenced by sensitive attributes such as user race, and pre-processing methods aim to mitigate such effects by adjusting certain item or user features to ensure fairness and diversity.

Pre-processing methods are typically simple, easy to integrate with existing IR systems and offer good generalizability. However, these methods are independent of the model and may remove certain features that are useful for the model.

\noindent\textbf{Model training.}
After preparing the training data, we first transform the raw data into vectorized representations (\ie embeddings) suitable for model input. Then, we design the IR models, assign appropriate loss functions, and optimize the models based on the defined loss functions.

The in-processing methods are mainly applied to the model training phase. Typically, they incorporate a fairness- and diversity-aware constraint or regularizer into the IR loss function, optimizing it to enhance ranking accuracy while ensuring fairness or diversity in the results~\cite{APR, FairNeg}. 

\noindent\textbf{Result evaluation.}
Finally, after training the Information Retrieval (IR) models, we apply them to evaluate their performance. We use the trained IR model to infer relevance scores for all user-item pairs in the test set. Based on these scores, we  generate a ranked list by selecting items with the highest relevance scores for each user.

Post-processing methods are often based on a given set of relevance scores and re-rank the items to form a new ranked list. They formulate the problem as a constrained linear programming optimization~\cite{fairrec, xu2023p, TaxRank}. The objective is to maximize the sum of relevance scores while fairness- and diversity-aware constraints will be incorporated to ensure a fair and diverse ranked list. 

Typically, post-processing methods are considered the most effective approach, but their performance is often impacted by errors propagated from earlier stages in the pipeline.

%% file: sections/details.tex
\section{Toolkit Details}

For package details, we introduce the datasets used, the implemented models, and the evaluation metrics adopted.

\subsection{Datasets}\label{sec:datasets}
We provide details of each used dataset of recommendation and search tasks in the following parts.

\begin{figure*}[t]  
    \centering    
    \includegraphics[width=\linewidth]{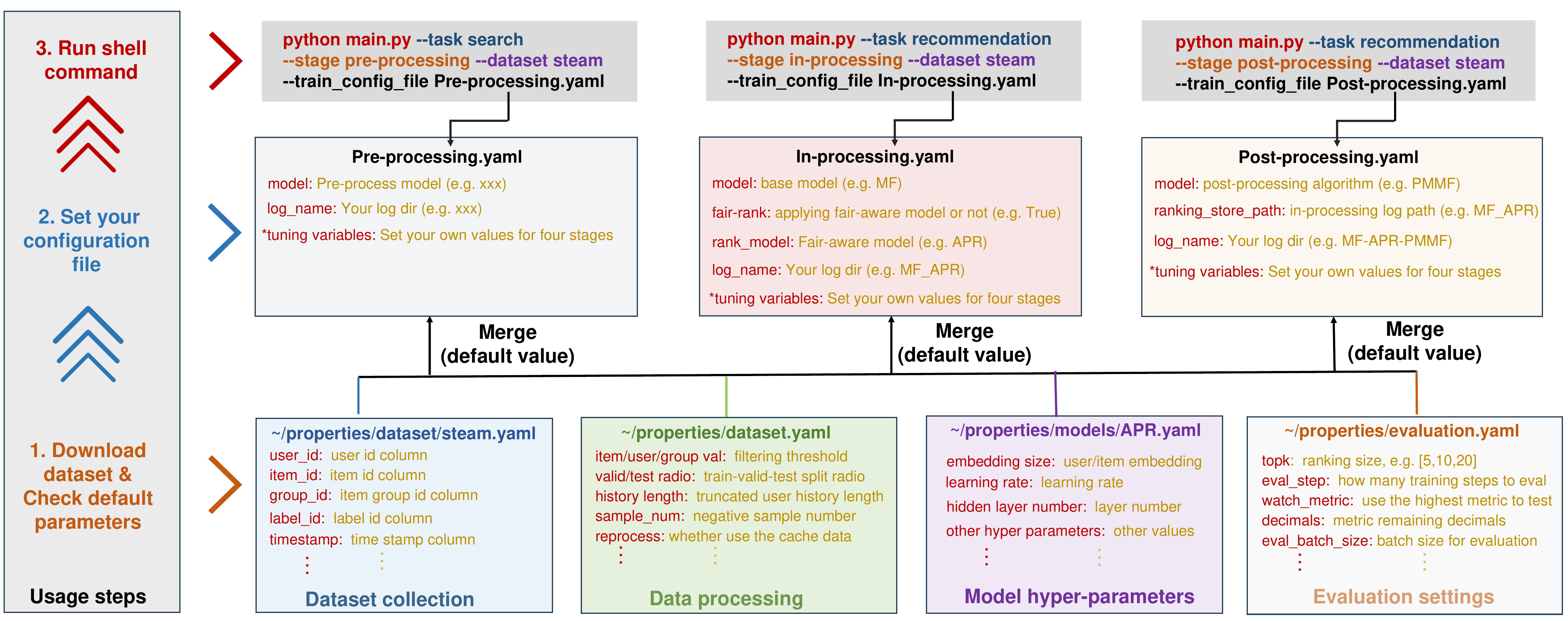}
    \caption{The usage of FairDiverse with three steps: (1) Download the datasets and check the default parameters of the four stages of pipelines; (2) Set custom configuration file to execute the pipeline. The $*$tuning\_variables allow you to define variable values for the default settings across the four pipeline stages, with the In-processing configuration file overriding these default values when specified; (3) Run the shell command, with the task, stage, dataset, and your custom configuration file. }
    \label{fig:recommendation_yamls}
\end{figure*}


\subsubsection{Recommendation} 
Any recommendation dataset can be used for the recommendation task. Specifically, we use the RecBole~\cite{recbole} dataset,\footnote{\url{https://github.com/RUCAIBox/RecSysDatasets}} which includes 43 commonly used datasets, all fully supported by our toolkit FairDiverse. The datasets span more than ten diverse domains, including games, products, and music. FairDiverse offers a comprehensive and fair comparison across all datasets.

To use them, researchers can simply download the datasets, place them in the \verb#~/recommendation/dataset# directory, and configure the settings in \verb#/properties/dataset/{dataset_name}.yaml#. The configuration files should specify which column names correspond to user ID, item ID, group ID, and other relevant fields. Once set up, running the command will enable algorithm evaluation on different datasets.

\subsubsection{Search.} 
Any dataset can be used with the pre-processing fairness models, however, in this framework we provide a working example on the COMPAS dataset \cite{bias2016there}. This dataset evaluates the bias in the COMPAS risk assessment tool by Northpointe (now Equivant), which predicts recidivism. It includes criminal history, recidivism probability, and sensitive attributes (gender and race). Following \cite{yang2020causal}, we provide a subset of 4,162 individuals distributed as 25\% White males, 59\% Black males, 6\% White females, and 10\% Black females.

For post-processing settings, we use ClueWeb09 Category B data collection~\cite{clueweb09_09_data} for our experiments.\footnote{\url{https://lemurproject.org/clueweb09.php/}} The ClueWeb09 dataset consists of 200 queries and 40,537 unique documents from the Web Track 2009-2012 dataset. Notably, queries \#95 and \#100 are not included in our experiments due to the lack of diversity judgments. The remaining 198 queries are associated with 3 to 8 manually annotated user intents, each accompanied by binary relevance ratings assigned at the intent level. 

\subsection{Models}\label{sec:models}
We provide all the implemented models in Table~\ref{tab:models}. Then we will delve into the details of each model. Note that to integrate different models into our toolkit architecture, some models may be re-implemented. As a result, their performance may vary due to differences in implementation and experimental settings.

\subsubsection{Recommendation.}
In the recommendation task, many models focus on in-processing and post-processing methods.

Firstly, we will categorize the base recommendation models into non-LLMs (Large Language Models) and LLMs-based models.
non-LLMs-based models mainly utilize user-item interaction behaviors to learn a good representation of users and items.
LLMs-based models rely on the prompts to rank the items according to their textual information such as item titles~\cite{LLMs_Fairness_Survey, DaiUncovering}. The  models are:
\begin{itemize}[leftmargin=*]
    \item Non-LLMs-based models: 
    \begin{itemize}[leftmargin=*]
        \item  \textbf{DMF}~\cite{DMF}: which optimizes the matrix factorization with the deep neural networks.
        \item \textbf{BPR}~\cite{BPR}: optimizes pairwise ranking via implicit feedback. \item \textbf{GRU4Rec}~\cite{tan2016improved}: employs gated recurrent units (GRUs) for session-based recommendations. \item \textbf{SASRec}~\cite{SASRec}: leverages self-attention mechanisms to model sequential user behavior.
    \end{itemize}
    
    \item LLMs-based models: \textbf{Llama3}~\cite{dubey2024llama}, \textbf{Qwen2}~\cite{bai2023qwen}, \textbf{Mistral}~\cite{jiang2023mistral7b}: utilizing rank-specific prompts to conduct ranking tasks under LLMs~\cite{DaiUncovering}.
\end{itemize}

\noindent%
We categorize in-processing models into re-weight, re-sample, regularizer, and prompt-based methods. 
The re-weighting and re-sample-based method adjusts sample weights/ratios during loss calculation, assigning higher weights/radios to underperforming item groups to enhance their support. Regularizer-based methods incorporate fairness- and diversity-aware regularization terms into the original loss function. In contrast, prompt-based methods, designed for LLM-based models, introduce fairness-aware prompts to enhance support for underperforming item groups. They are:

\begin{itemize}[leftmargin=*]
    \item Re-weight-based models: 
    \begin{itemize}[leftmargin=*]
        \item \textbf{APR}~\cite{APR}: an adaptive reweighing method that dynamically prioritizes samples near the decision boundary to mitigate distribution shifts.
        \item \textbf{FairDual}~\cite{FairDual}: applies dual-mirror gradient descent to dynamically compute the weight for each sample to support the worst-off groups.
        \item \textbf{IPS}~\cite{jiang2024item}: employs the reciprocal of the sum popularity of items within the group as the weight assigned to that group.
        \item \textbf{Minmax-SGD}~\cite{Minmax-SGD}: applies optimizing techniques to dynamically sample groups.
        \item \textbf{SDRO}~\cite{SDRO}: Improves DRO with the distributional shift to optimize group MMF.
    \end{itemize}
   
    \item Re-sample-based models: \textbf{FairNeg}~\cite{FairNeg}: adjusts the group-level negative sampling distribution in the training process.
    \item Regularizer-based models: 
    \begin{itemize}[leftmargin=*]
        \item \textbf{DPR}~\cite{DPR}: applies a fair-aware adversarial loss based on statistical parity and equal opportunity.
        \item \textbf{FOCF}~\cite{yao2017beyond}: applies a fair-aware regularization loss of different groups.
        \item \textbf{Reg}~\cite{Reg}: applies a penalty on the squared difference between the scores of two groups across all positive user-item pairs.
    \end{itemize}
    
    \item Prompt-based models: \textbf{FairPrompts}~\cite{xu-etal-2024-study}.\footnote{We do not fine-tune the LLMs but only use manually designed prompts.}
\end{itemize}

\noindent%
We categorize the post-processing models into heuristic and learning-based models. Heuristic models primarily use algorithms like greedy search to re-rank items, while learning-based models dynamically generate fairness- and diversity-aware scores, which are incorporated into the original relevance score for re-ranking. They are:


\begin{itemize}[leftmargin=*]
    \item Heuristic models: 
    \begin{itemize}[leftmargin=*]
        \item \textbf{CP-Fair}~\cite{cpfair}: applies a greedy solution to optimize the knapsack problem of fair ranking.
        \item \textbf{min-regularizer}~\cite{xu2023p}: adds an additional fairness score to the ranking scores, capturing the gap between the current utility and the worst-off utility.
        \item \textbf{RAIF}~\cite{liu2025repeat}: a model-agnostic repeat-bias-aware item fairness optimization algorithm based on mixed-integer linear programming.\footnote{Note that we remove repeat bias term, change the item fairness objective to make the exposure of each group closer, and extend RAIF into multi-group cases.}
    \end{itemize}
    
    \item Learning-based methods: 
    \begin{itemize}[leftmargin=*]
        \item \textbf{P-MMF}~\cite{xu2023p}: applies a dual-mirror gradient descent method to optimize the accuracy-fairness trade-off problem.
        \item \textbf{FairRec}~\cite{fairrec}, \textbf{FairRec+}~\cite{fairrecplus}: proposes leveraging Nash equilibrium to guarantee Max-Min Share of item exposure.
        \item \textbf{FairSync}~\cite{fairsync}: proposes to guarantee the minimum group utility under distributed retrieval stages.
        \item \textbf{Tax-Rank}~\cite{TaxRank}: applies the optimal transportation (OT) algorithm to trade-off fairness-accuracy.
        \item \textbf{Welf}~\cite{nips21welf}: use the Frank-Wolfe algorithm to maximize the Welfare functions of worst-off items.
    \end{itemize}
    
\end{itemize}

\subsubsection{Search.}
Our framework makes use of the Ranklib library to offer a variety of ranking models, including 8 popular algorithms: MART \cite{friedman2001greedy}, RankNet \cite{burges2005learning}, RankBoost \cite{freund2003efficient}, AdaRank \cite{xu2007adarank}, Coordinate Ascent \cite{metzler2007linear}, LambdaMART \cite{wu2010adapting}, ListNet \cite{cao2007learning}, Random Forests \cite{breiman2001random}. 

We divide pre-processing models into two categories causal based models and probabilistic mapping clustering models. All model implementations are adapted to optimize for multiple sensitive attributes and for non-binary groups, including intersectional groups.
\begin{itemize}[leftmargin=*]
    \item Causal based models:
    \begin{itemize}[leftmargin=*]
        \item \textbf{CIF-Rank}~\cite{yang2020causal} estimates the causal effect of the sensitive attributes on the data and makes use of them to correct for the bias encoded.
    \end{itemize}
    \item Probabilistic mapping clustering models:
    create representations which are independent of the available sensitive attributes.
    \begin{itemize}[leftmargin=*]
        \item \textbf{LFR}~\cite{zemel2013learning} optimizes for group fairness by making sure that the probability of a group to be mapped to a cluster is equal to the probability of the other group. 
        \item \textbf{iFair}~\cite{lahoti2019ifair} optimizes for individual fairness by making sure that the distance between similar individuals is maintained in the new space. 
        \item \textbf{gFair} based on iFair~\cite{lahoti2019ifair}, optimizes for group fairness by making sure that the distance between similar individuals from a group are close to similar individuals from the other group. It constraints the optimization to maintain the relative distance between individuals belonging to the same group.
    \end{itemize}
\end{itemize}
        
\noindent%
For post-processing search models, we often utilize the diversity-aware re-ranking models. These models can be roughly categorized into unsupervised methods and supervised methods. 

\begin{itemize}[leftmargin=*]
    \item Unsupervised methods:
    \begin{itemize}[leftmargin=*]
        \item  \textbf{PM2}~\cite{PM2_12_sigir}: optimizes proportionality by iteratively determining the topic that best maintained the overall proportionality. 
        \item \textbf{xQuAD}~\cite{xQuAD_10_www}: utilizes sub-queries representing pseudo user intents and diversifies document rankings by directly estimating the relevance of the retrieved documents to each sub-queries. 
        \item \textbf{DiversePrompts}: a diversity ranking model based on large language models. We design specific prompts tailored for search result diversification based on two latest closed-source LLMs: GPT-4o~\cite{openai2024gpt4technicalreport} and Claude 3.5~\cite{Claude3.5}.
    \end{itemize}
    \item Supervised methods: 
    \begin{itemize}[leftmargin=*]
        \item \textbf{DESA}~\cite{DESA_20_cikm}: employs the attention mechanism to model the novelty of documents and the explicit subtopics. 
        \item \textbf{DALETOR}~\cite{DALETOR_21_WWW}: proposes diversification-aware losses to approach the optimal ranking. 
    \end{itemize}
\end{itemize}

\subsection{Evaluation Metrics}\label{sec:evaluation}
We will delve into the details of each used evaluation metrics in the following parts.

\noindent\textbf{Recommendation.} In recommendation, evaluation metrics are generally categorized into two types: 
\begin{itemize}[leftmargin=*]
    \item Ranking accuracy-based metric: Mean Reciprocal Rank (MRR), Hit Radio (HR), and Normalized Discounted Cumulative Gain (NDCG)~\cite{IRbook}, utility loss (\ie Regret)~\cite{xu2023p}~\footnote{Note that evaluation metric NDCG in post-processing is slightly different compared to common definition: NDCG in post-processing means the re-ranking quality compared to original ranking quality~\cite{xu2023p}.}.
    \item Fairness- and diversity-based metric:  MMF~\cite{xu2023p}, GINI index~\cite{nips21welf}, Entropy~\cite{jost2006entropy}, and MinMaxRatio~\cite{rehman2018selection}.
\end{itemize}

\noindent\textbf{Search.} For the search task, we adopt the official diversity evaluation metrics of the Web Track, including ERR-IA~\cite{erria_09_cikm}, $\alpha$-nDCG~\cite{andcg_08_sigir}, and the diversity measure Subtopic Recall (denoted as S-rec)~\cite{srec15sigir}. These metrics assess the diversity of document rankings by explicitly rewarding novelty while penalizing redundancy. We follow the Web Track and utilize the provided shell command to evaluate model performance.

Furthermore, we provide support for fairness metrics, including group fairness measures such as demographic parity, ensuring proportional representation of groups, as well as proportional exposure of groups. Additionally, one can compute the in-group fairness metric proposed by \citet{yang2019balanced}, which computes the ratio between the lowest accepted score and the highest rejected score within a group. On top of group fairness, one can compute individual fairness by doing a pairwise comparison between candidates' distance in the features space and their achieved exposure~\cite{dwork2012fairness}. 

%% file: sections/usage.tex
\section{Toolkit Usage}


Figure~\ref{fig:recommendation_yamls} provides an overview of the three main steps for utilizing our toolkit, FairDiverse. We will describe each step of usage in detail. The detailed configuration file parameters can be found in~\url{https://xuchen0427.github.io/FairDiverse/}.

\subsection{Usage Steps}
\textbf{Step 1.} First, download the dataset you wish to test, as described in Section~\ref{sec:datasets}, and store it in the \texttt{/dataset} directory. Then, specify the parameters in \texttt{/properties/dataset/\{data\_name\}.yaml.}
Next, you need to review all default parameters for data processing, model hyperparameters, and evaluation settings.

\noindent\textbf{Step 2.} Then, you need to create your own configuration file, specifying the selected models and the log directory path. If you want to modify the default pipeline parameters, you can specify them directly in your own configuration file, which will override the values in the default configuration files.

\noindent\textbf{Step 3.} Enter \texttt{/fairdiverse} dictionary and execute the command

\begin{lstlisting}[style=shell]
python main.py --task "recommendation" 
--stage "in-processing" --dataset "steam" 
--train_config_file "In-processing.yaml"
\end{lstlisting}

\noindent%
The args should specify the task (recommendation/search), stage (pre-processing, in-processing, post-processing), dataset, and your custom configuration file as defined in Step 2. Finally, the evaluation results and item/user utility allocations will then be recorded in your specified log file.

\subsection{Usage Example}
Besides our provided \texttt{main.py} file and shell command, you can also utilize the following test codes to run the toolkit.

\noindent\textbf{Recommendation.}
Our repository includes an example dataset, Steam.\footnote{\url{http://cseweb.ucsd.edu/~wckang/Steam_games.json.gz}} We provide the simple code snippets for running the in-processing and post-processing models, which are listed below. The user needs to specify the chooses ``model,'' ``dataset'' and ``log\_name'' for training and testing. 

\begin{lstlisting}[language=Python]
from recommendation.trainer import RecTrainer

config = {'model': 'BPR', 'data_type': 'pair', 'fair-rank': True, 'rank_model': 'APR', 'use_llm': False, 'log_name': "test", 'dataset': 'steam'}

trainer = RecTrainer(train_config=config)
trainer.train()
\end{lstlisting}

\begin{lstlisting}[language=Python]
from recommendation.reranker import RecReRanker

config = {'ranking_store_path': 'steam-base-mf', 'model': 'CPFair', 'fair-rank': True, 'log_name': 'test', 'fairness_metrics': ["GINI"], 'dataset': 'steam'}

reranker = RecReRanker(train_config=config)
reranker.rerank()
\end{lstlisting}

\noindent\textbf{Search.}
Our repository includes a running example of the pre-processing models on the COMPAS dataset. We provide the simple code snippet for running the pre-processing models, listed as follows. One can set the ``preprocessing\_model'' field to any of the supported models: CIFRank, LFR, gFair and iFair. Each pre-processing model has its own config file under \texttt{search/properties/models}  which is automatically loaded based on your choice. 

\begin{lstlisting}[language=Python]
from search.trainer_preprocessing_ranker import RankerTrainer
 
config={"train_ranker_config": {"preprocessing_model": "iFair", "name": "Ranklib", "ranker": "RankNet", "lr": 0.0001, "epochs": 10}}
 
reranker=RankerTrainer(train_config=config)
reranker.train()
\end{lstlisting}

For the post-processing models, our repository also provides a running example on the ClueWeb09 dataset. The simple code snippet for running these models is shown as follows. 

\begin{lstlisting}[language=Python]
from search.trainer import SRDTrainer
   
config={'model':'xquad', 'dataset':'clueweb09', 'log_name': 'test', 'model_save_dir': "model/", 'tmp_dir': "tmp/", 'mode': "train",}
 
trainer = SRDTrainer(train_config=config)
trainer.train()
\end{lstlisting}


%% file: sections/experiments.tex
\begin{table*}[t]
\caption{Partial benchmark results for recommendation tasks on Steam datasets for different ranking sizes $K$. They are evaluated on BPR ranking models with in-processing fairness-aware and diversity-aware approaches. $\downarrow$ and $\uparrow$ indicate that a smaller or larger metric value, respectively, corresponds to better model performance. It is important to note that the reported results are based on default parameters. }
\label{tab:exp:rec_in_processing}
\small
\setlength{\tabcolsep}{1.2mm}
\begin{tabular}{l cccccc cccccc}
\toprule
\multicolumn{1}{@{}l}{\multirow{2}{*}{Models/Metric}} & \multicolumn{6}{c}{$K=10$} & \multicolumn{6}{c}{$K=20$} \\
\cmidrule(r){2-7}
\cmidrule{8-13}
& NDCG$\uparrow$ & MRR$\uparrow$ & HR$\uparrow$ & MMF$\uparrow$ & GINI$\downarrow$ & Entropy$\uparrow$  & NDCG$\uparrow$ & MRR$\uparrow$ & HR$\uparrow$ & MMF$\uparrow$ & GINI$\downarrow$ & Entropy$\uparrow$ \\ 
\midrule
Llama3-FairPrompts  & 0.0304 & 0.0565 & 0.0265 & 0.0364 & 0.7332 & 3.8800  & 0.0444 & 0.1083 & 0.0308 & 0.0980 & 0.6201 & 4.3738  \\
Qwen2-FairPrompts   & 0.0312 & 0.0546 & 0.0284 & 0.0324 & 0.7503 & 3.7629  & 0.0455 & 0.1070 & 0.0329 & 0.0871 & 0.6453 & 4.2590  \\
Mistral-FairPrompts & 0.0323 & 0.0559 & 0.0303 & 0.0315 & 0.7494 & 3.7751  & 0.0481 & 0.1132 & 0.0355 & 0.0861 & 0.6455 & 4.2602  \\
APR  & 0.2925 & 0.2934 & 0.4085 & 0.0324 & 0.7257 & 3.9513  & 0.3193 & 0.2999 & 0.5058 & 0.0590 & 0.6485 & 4.2997  \\
FairDual    & 0.3204 & 0.3073 & 0.4727 & 0.0330 & 0.7123 & 4.0301  & 0.3479 & 0.3136 & 0.5702 & 0.0563 & 0.6577 & 4.2745  \\
IPS  & 0.3073 & 0.3090 & 0.4213 & 0.0249 & 0.7314 & 3.9183  & 0.3347 & 0.3155 & 0.5196 & 0.0570 & 0.6517 & 4.2824  \\
Minmax-SGD  & 0.2672 & 0.2604 & 0.3961 & 0.0218 & 0.7501 & 3.7252  & 0.2958 & 0.2679 & 0.4991 & 0.0470 & 0.6936 & 3.9982  \\
SDRO & 0.3009 & 0.3056 & 0.4081 & 0.0350 & 0.7212 & 3.9754  & 0.3298 & 0.3124 & 0.5137 & 0.0619 & 0.6451 & 4.3156  \\
FairNeg     & 0.2964 & 0.2975 & 0.4125 & 0.0673 & 0.6671 & 4.2222  & 0.3208 & 0.3036 & 0.5020 & 0.0778 & 0.6158 & 4.4067  \\
FOCF & 0.2879 & 0.2879 & 0.4041 & 0.0294 & 0.7272 & 3.9460  & 0.3141 & 0.2942 & 0.4992 & 0.0579 & 0.6472 & 4.3086  \\
Reg  & 0.2979 & 0.2981 & 0.4162 & 0.0306 & 0.7270 & 3.9465  & 0.3245 & 0.3043 & 0.5127 & 0.0584 & 0.6497 & 4.2917  \\ 
\bottomrule
\end{tabular}
\end{table*}

\begin{table*}[t]
 \caption{Partial benchmark results for recommendation tasks on ClueWeb datasets for different ranking sizes $K$. They can be evaluated using the shell command provided in our GitHub repository. It is important to note that the reported results are based on default parameters.}
\label{tab:exp:rec_post_processing}
\small
\setlength{\tabcolsep}{1.2mm}
\begin{tabular}{l cccccc cccccc}
\toprule
\multicolumn{1}{@{}l}{\multirow{2}{*}{Models/Metric}} & \multicolumn{6}{c}{$K=10$} & \multicolumn{6}{c}{$K=20$} \\
\cmidrule(r){2-7}
\cmidrule{8-13}
\multicolumn{1}{@{}l}{} & R-NDCG$\uparrow$ & u-loss$\downarrow$ & MMF$\uparrow$ & GINI$\downarrow$ & Entropy$\uparrow$ & MMR$\uparrow$  & R-NDCG$\uparrow$ & u-loss$\downarrow$ & MMF$\uparrow$ & GINI$\downarrow$ & Entropy$\uparrow$ & MMR$\uparrow$ \\
\midrule
CP-Fair & 0.9981 & 0.0035 & 0.2135 & 0.4424 & 4.8441 & 0.0196 & 0.9969 & 0.0055 & 0.2118 & 0.4349 & 4.9064 & 0.0301 \\
min-regularizer & 0.9272 & 0.1234 & 0.3984 & 0.1373 & 5.3796 & 0.2740 & 0.9359 & 0.0896 & 0.4004 & 0.1326 & 5.3818 & 0.2761 \\
RAIF & 0.9881 & 0.0233 & 0.2937 & 0.3293 & 5.0080 & 0.0248 & 0.9829 & 0.0302 & 0.3333 & 0.2585 & 5.1558 & 0.0358 \\
P-MMF & 0.9691 & 0.0536 & 0.3140 & 0.2792 & 5.1911 & 0.0685 & 0.9675 & 0.0482 & 0.3289 & 0.2429 & 5.2597 & 0.0987 \\
FairRec & 0.9529 & 0.1088 & 0.1866 & 0.5098 & 4.5372 & 0.0157 & 0.9497 & 0.0990 & 0.1779 & 0.5179 & 4.5234 & 0.0175 \\
FairRec+ & 0.9773 & 0.0540 & 0.1758 & 0.5254 & 4.5108 & 0.0119 & 0.9750 & 0.0510 & 0.1609 & 0.5463 & 4.4594 & 0.0134 \\
FairSync & 0.9816 & 0.0353 & 0.2477 & 0.3951 & 4.9152 & 0.0237 & 0.9785 & 0.0383 & 0.2553 & 0.3705 & 5.0165 & 0.0312 \\
TaxRank & 0.9438 & 0.1015 & 0.2421 & 0.3791 & 4.9846 & 0.0149 & 0.9303 & 0.1080 & 0.2907 & 0.3078 & 5.1287 & 0.0209 \\
Welf & 0.9668 & 0.0575 & 0.3216 & 0.2638 & 5.2254 & 0.0820 & 0.9682 & 0.0467 & 0.3322 & 0.2383 & 5.2674 & 0.1112 \\ \bottomrule
\end{tabular}
\end{table*}

\section{Benchmark Results Analysis}
In this section, we present an analysis of partial benchmark results derived from our toolkit FairDiverse. Note that our goal is not to compare different models but to highlight the analytical direction of these results, helping researchers interpret and understand the findings more effectively and efficiently.

\subsection{Recommendation} 

In recommendation, we primarily evaluate the performance of commonly used in-processing models, which integrate fairness constraints during training, and post-processing models, which adjust rankings after predictions to enhance fairness.

\noindent\textbf{In-processing models.}
Table~\ref{tab:exp:rec_in_processing} presents the performance of the implemented in-process fairness and diversity-aware models in terms of accuracy (NDCG, HR, MRR) and fairness/diversity (MMF, GINI, Entropy) under the default parameters of our toolkit. The benchmark results are obtained using the Steam dataset and BPR~\cite{BPR} ranking models. The fairness/diversity metric is calculated at the Steam game category level.

First, from Table~\ref{tab:exp:rec_in_processing}, we observe that LLM-based models generally exhibit higher fairness and diversity but lower ranking performance. In contrast, non-LLM-based models achieve better ranking performance but struggle with the long-tail problem. 
Secondly, different methods often exhibit significant variance in accuracy and fairness/diversity performance, excelling in some metrics while underperforming in others. Moreover, the trends across different fairness metrics are not always consistent. 

Our toolkit provides researchers with a unified and convenient tool to compare various methods, explore trade-offs between different metrics, and analyze the reasons behind performance gaps. Researchers can use our toolkit to analyze results and validate their ideas across different base models and datasets.

\noindent\textbf{Post-processing models.} Table~\ref{tab:exp:rec_post_processing} shows the results of implemented post-process fairness and diversity-aware models in terms of re-ranking accuracy (NDCG, u-loss) and fairness/diversity (MMF, GINI, Entropy, and MMR). The benchmark results are obtained using the Steam dataset and the ranking lists provided from DMF~\cite{DMF} models. The fairness/diversity metric is also calculated at the Steam game category level.

First, from Table~\ref{tab:exp:rec_in_processing}, we observe that post-processing models outperform in-processing methods in fairness and diversity. However, they often face an accuracy-fairness trade-off, sacrificing accuracy to enhance the fairness and diversity of item categories. With our toolkit, FairDiverse, researchers can explore this trade-off across different parameters, base models, and datasets.

\begin{figure*}[t]  
    \centering    
    \includegraphics[width=\linewidth]{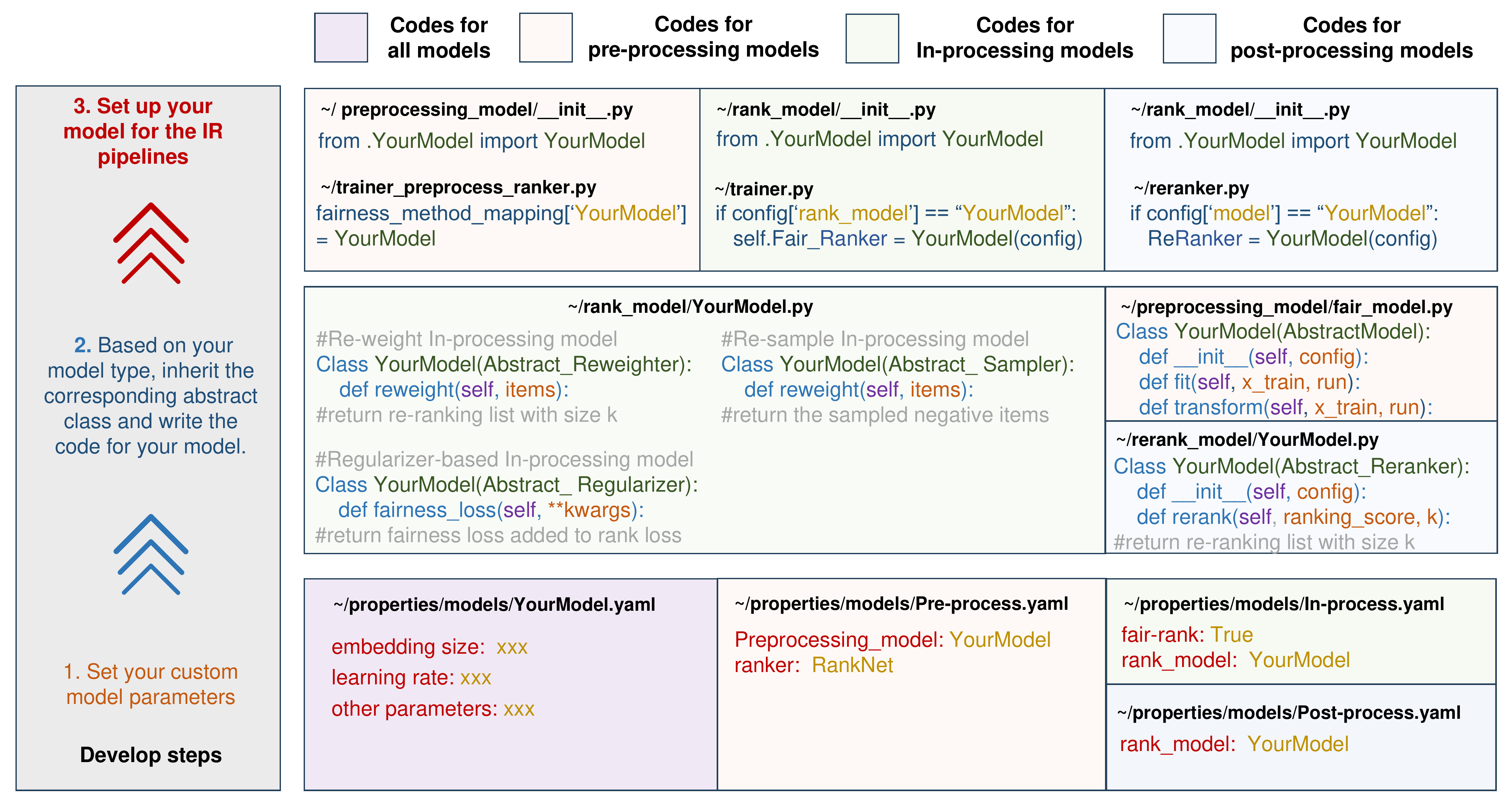}
    \caption{The custom steps for fairness and diversity-aware search and recommender models named \textit{YourModel}. The differently colored areas indicate the code you need to add when developing  different types of model. Generally, you can follow three steps: (1) define custom model parameters, (2) develop your model based on its type, and (3) integrate it into the pipeline. }
    \label{fig:rec_APIs}
\end{figure*}


\subsection{Search}
\begin{table*}[t]
 \caption{Benchmark results for search task on the COMPAS dataset for different ranking sizes $K$, obtained on RankNet ranking models with pre-processing fairness-aware approaches. Evaluated using the shell command provided in our GitHub repository. \%D: diversity of the Female-Black group (the disadvantaged intersectional group); IGF: in-group-fairness measure as an average over the groups; yNN: individual fairness; NDCG-loss: NDCG loss. The reported results are based on default parameters.}
\label{tab:exp:rec_pre_processing}
\small
\setlength{\tabcolsep}{1.3mm}
\begin{tabular}{l cccc cccc }
\toprule
\multicolumn{1}{@{}l}{\multirow{2}{*}{Models/Metric}} & \multicolumn{4}{c}{$K=100$}      & \multicolumn{4}{c}{$K=300$} \\
\cmidrule(r){2-5}
\cmidrule{6-9}
\multicolumn{1}{@{}l}{} & \%D$\uparrow$ & IGF$\uparrow$ & yNN$\uparrow$ & NDCG-loss$\uparrow$ & \%D$\uparrow$ & IGF$\uparrow$ & yNN$\uparrow$ & NDCG-loss$\uparrow$ \\ 
\midrule
RankNet & 0.10 & - & 0.86 & 0.92 & 0.11 & - & 0.86 & 0.95  \\
CIFRank & 0.13 & 1.00 & 0.86 & 0.78 & 0.10 & 1.00 & 0.86 & 0.84 \\
LFR & 0.09 & 1.00 & 0.86 & 0.93 & 0.09 & 0.93 & 0.86 & 0.72 \\
gFair & 0.14 & 1.00 & 0.86 & 0.39 & 0.10 & 1.00 & 0.86 & 0.52 \\
iFair & 0.42 & 0.55 & 0.86 & 0.85 & 0.19 & 0.30 & 0.86 & 0.89 \\ \bottomrule
\end{tabular}
\end{table*}

\begin{table*}[t]
 \caption{Benchmark results for the post-processing search result diversification models on the ClueWeb09 datasets with different ranking sizes $K$. We evaluate the performance using the shell command provided by the official Web Track which is also available in our GitHub repository. A larger metric value indicates superior model performance. The reported results are based on default parameter settings. }
\label{tab:exp:search_post_processing}
\small
\setlength{\tabcolsep}{1.2mm}
\begin{tabular}{l ccc ccc ccc}
\toprule
\multicolumn{1}{@{}l}{\multirow{2}{*}{Models/Metric}} & \multicolumn{3}{c}{$K=5$} & \multicolumn{3}{c}{$K=10$} & \multicolumn{3}{c}{$K=20$} \\
\cmidrule(r){2-4}
\cmidrule(r){5-7}
\cmidrule{8-10}
\multicolumn{1}{@{}l}{} & ERR-IA & $\alpha$-nDCG & S-rec & ERR-IA & $\alpha$-nDCG & S-rec & ERR-IA & $\alpha$-nDCG & S-rec \\ 
\midrule
PM2 & 0.2626 & 0.3292 & 0.4793 & 0.2824 & 0.3684 & 0.5708 & 0.2913 & 0.3989 & 0.6407 \\
xQuAD & 0.2002 & 0.2511 & 0.3961 & 0.2166 & 0.2838 & 0.4701 & 0.2272 & 0.3230 & 0.5761\\
DiversePrompts (GPT-4o) & 0.2890 & 0.3514 & 0.4972 & 0.3054 & 0.3833 & 0.5791 & 0.3131 & 0.4099 & 0.6396\\
DiversePrompts (Claude 3.5) & 0.3136 & 0.3800 & 0.4981 & 0.3292 & 0.4079 & 0.5741 & 0.3372 & 0.4348 & 0.6486 \\
DESA & 0.3497 & 0.4226 & 0.5195 & 0.3642 & 0.4452 & 0.5914 & 0.3703 & 0.4655 & 0.6438\\
DALETOR & 0.2770 & 0.3362 & 0.4609 & 0.2948 & 0.3732 & 0.5644 & 0.3047 & 0.4085 & 0.6581\\
\bottomrule
\end{tabular}
\end{table*}

As for the search task, we first evaluate the output of a ranking model trained on data that was debiased by a pre-processing model and then observe the diversity of the final ranking results achieved by post-processing models.


\noindent\textbf{Pre-processing models.} The results of implemented pre-pro\-cessing models applied on the input of a ranking model are denoted in Table~\ref{tab:exp:rec_pre_processing}. In this setting the ranking model is RankNet using the implementation provided by the Ranklib Library. We evaluate the performance on the COMPAS dataset. NDCG-loss represents the loss in utility with respect to the original ranking and the original scores. 

All models, except LFR, manage to improve or maintain the diversity in top-k of the Female-Black group, which is the most disadvantaged intersectional group. Out of all models gFair has the biggest loss in utility, while individual fairness (yNN) is not affected. In-group-fairness (IGF) is measured on the transformed representations, not on the output ranking, to check whether the transformed data respects the order within a group. It can be observed that CIFRank and gFair obtain perfect IGF. Using FairDiverse researchers can compare the impact of pre-processing models on the output ranking given the trade-offs between group fairness, individual fairness and utility loss. 

\noindent\textbf{Post-processing models.} The results of implemented post-process search result diversification models are denoted in Table~\ref{tab:exp:search_post_processing}. We evaluate the performance based on the ClueWeb09 dataset. The initial ranking list is provided by Lemur.\footnote{\url{https://lemurproject.org/clueweb09.php/}} We utilize the top 50 documents from the initial ranking list for testing these diversified ranking models' performance.

From the results, we can observe that, first, supervised diversified search models demonstrate superior performance compared to unsupervised models. Moreover, diversified rankers based on LLMs consistently outperform traditional unsupervised methods. This observation suggests that the knowledge acquired by LLMs during the pre-training stage significantly enhances the diversity of search results. To facilitate the exploration of these models, we present FairDiverse, a comprehensive toolkit that enables researchers to analyze various parameters, base models, and datasets.

%% file: sections/APIs.tex
\section{Customizing Models in FairDiverse}
We outline the steps for customizing and evaluating new IR models using the APIs we provide. Detailed API descriptions and source code can be found in~\url{https://xuchen0427.github.io/FairDiverse/}. 
The provided APIs can be used by installing them via pip: 

\begin{lstlisting}[style=shell]
pip install fairdiverse
\end{lstlisting}


\subsection{Steps}
Figure~\ref{fig:rec_APIs} illustrates the three key steps for implementing fairness- and diversity-aware IR models named \textit{YourModel}.

\noindent\textbf{Step 1.} Configure your custom model parameters and save them in a newly created \texttt{YourModel.yaml} file in the \texttt{/properties/models/} directory. Then you can change the model in the running configuration file \texttt{Post-processing.yaml}.

\noindent\textbf{Step 2.} Select the appropriate Python abstract class from our provided options based on your model type and implement your model in a newly created file, \texttt{YourModel.py}, stored in the corresponding directory. You can use the integrated tools and common parameters within the abstract class. Researchers only need to focus on designing the model without worrying about the rest of the pipeline. 


\noindent\textbf{Step 3.} Configure your model for the training pipeline by following these steps: import your custom model package in the corresponding file (\texttt{/model\_type/\_\_init\_\_.py}) and define the model in the appropriate script (\texttt{/train.py}, \texttt{/reranker.py}).


\subsection{Examples}
Here, we provide two example codes demonstrating how to design a custom search and recommendation model, respectively. 
\begin{lstlisting}[language=Python]
#/recommendation/rank_model/YourModel.py
class YourModel(Abstract_Regularizer):
    def __init__(self, config, group_weight):
        super().__init__(config)

    def fairness_loss(self, input_dict):
        return torch.var(input_dict['scores'])

#/recommendation/rank_model/__init__.py
from .YourModel import YourModel

#/recommendation/trainer.py
if config["model"] == "YourModel":
  self.Model = YourModel(config)
\end{lstlisting}

\begin{lstlisting}[language=Python]
#/search/preprocessing_model/YourModel.py
class YourModel(PreprocessingFairnessIntervention):
    def __init__(self, configs, dataset):
        super().__init__(configs, dataset)

    def fit(self, X_train, run):
    # Train the fairness model using the training set.

    def transform(self, X_train, run file_name=None):
    # Apply the fairness transformation to the dataset.

#/search/preprocessing_model/__init__.py
from .YourModel import YourModel
fairness_method_mapping['YourModel'] = YourModel

\end{lstlisting}

%% file: sections/related_work.tex
\section{Related Work}\label{sec:related_work}

\textbf{Beyond-Accuracy in IR.} In modern IR systems, beyond-accuracy objectives play a crucial role in building a more effective and responsible ecosystem~\cite{kaminskas2016diversity, de2023beyond}. Beyond-accuracy objectives primarily include diversity~\cite{PM2_12_sigir}, fairness~\cite{xu2023p, fairrec}, novelty~\cite{hurley2011novelty}, and serendipity~\cite{zhang2012auralist}. Among these factors, this toolkit primarily focuses on fairness and diversity.


\noindent\textbf{Fairness and diversity in IR.} Fairness and diversity are gaining increasing attention in the IR field, as both seek to support underrepresented user and item groups~\cite{li2022fairness, LLM4FairSurvey, wang2021user, PM2_12_sigir}. Previous studies have often explored fairness and diversity from the perspectives of different stakeholders, such as users and items~\cite{abdollahpouri2020multistakeholder}, as well as at varying granularities, including both group-level and individual-level fairness~\cite{biega2018equity, xu2023p}. Based on different stages in the IR pipeline, previous methods are often categorized into three types: pre-processing~\citep{rus2024study}, in-processing~\citep{APR, FairNeg, Reg}, and post-processing approaches~\cite{xu2023p, TaxRank, PM2_12_sigir, dang2012diversity}. As for the evaluation, they are also based on different metrics, including the Gini index~\cite{nips21welf}, MMF~\cite{xu2023p} in recommendation, and $\alpha$-nDCG~\cite{andcg_08_sigir}, NRBP~\cite{nrbp_09_ictir} in search. However, fairness and diversity often lack unified evaluation settings. This paper introduces FairDiverse, a benchmarking toolkit designed to comprehensively assess different models under different IR tasks.

\noindent\textbf{Fairness and diversity toolkits.} Most fairness and diversity toolkits are implemented under classification tasks. For example, FFB~\cite{han2023ffb} implements diverse in-processing models for addressing group fairness problems. Fairlearn\cite{bird2020fairlearn},  AIF360\cite{aif360-oct-2018} and Aequitas\cite{jesus2024aequitas} implement the unfairness mitigation algorithms using Scikit-learn~\cite{kramer2016scikit} API design. However, these methods cannot be directly applied to ranking tasks. Although some toolkits~\cite{recbole2.0} have been proposed to incorporate fairness and diversityd in IR, they primarily focus on recommendation tasks and implement only a limited number of in-processing models. Our toolkit, FairDiverse, offers the most extensive collection of models, covering a wide range of fairness- and diversity-aware algorithms across both search and recommendation tasks. Moreover, FairDiverse is highly extensible, offering flexible APIs for easy integration of new fairness- and diversity-aware IR models, unlike other toolkits with complex class inheritance.

%% file: sections/Conclusion.tex
\section{Conclusion}

We have presented FairDiverse, a comprehensive IR toolkit designed to facilitate standardized evaluations of fairness-aware and diversity-aware algorithms. FairDiverse offers three key advantages over other toolkits. (1) It provides a comprehensive framework for integrating various models at different IR pipeline stages; (2) It implements 28 fairness- and diversity-aware models for 16 recommendation and search base models; (3) It offers flexible APIs for researchers to develop and integrate custom models.

However, a limitation of FairDiverse is that it only supports single-round fairness- and diversity-aware IR models. It does not yet support dynamic settings, such as long-term fairness or fairness under dynamic feedback loops~\cite{xu2023ltp}. In future work, we plan to expand FairDiverse to support dynamic scenarios and explore the use of LLM agents~\cite{zhang2024generative} for simulation and evaluation. Additionally, we also aim to incorporate more beyond-accuracy models, including novelty- and serendipity-aware models.